\documentclass[preprint,
 superscriptaddress,
 bibnotes,
 amsmath,amssymb, aps,
 prd,
floatfix,
]{revtex4-1}

\usepackage{url}
\usepackage{soul}
\usepackage{float}
\usepackage[utf8]{inputenc}
\usepackage{graphicx}
\usepackage[caption=false]{subfig}
\usepackage{dcolumn}
\usepackage{multirow}
\usepackage{tabularx}
\usepackage{bm}
\usepackage{hyperref}
\usepackage{amsmath}
\usepackage{amssymb}
\usepackage{comment}
\usepackage{color}
\usepackage{enumerate}
\usepackage{booktabs}
\usepackage{siunitx}
\usepackage{soul}

\newcommand{\be}{\begin{eqnarray}}
\newcommand{\ee}{\end{eqnarray}}

\usepackage[usenames,dvipsnames,svgnames]{xcolor}

\begin{document}

\title{Magnetic Bubble Chambers and Sub-GeV Dark Matter Direct Detection}%

 \author{Philip C. Bunting}%
\email{pcb7445@berkeley.edu}
\affiliation{Department of Chemistry, University of California, Berkeley, California 94720, USA}

 \author{Giorgio Gratta}%
\email{gratta@stanford.edu}
\affiliation{Physics Department and HEPL, Stanford University, Stanford CA 94305, USA}

 \author{Tom Melia}%
\email{tmelia@lbl.gov}
\affiliation{Department of Physics, University of California, Berkeley, California 94720, USA}
\affiliation{Theoretical Physics Group, Lawrence Berkeley National Laboratory, Berkeley, California 94720, USA}
\affiliation{Kavli Institute for the Physics and Mathematics of the Universe (WPI), University of Tokyo, Kashiwa 277-8583, Japan}

 \author{Surjeet Rajendran}%
\email{surjeet@berkeley.edu}
\affiliation{Department of Physics, University of California, Berkeley, California 94720, USA}

\preprint{}%

\begin{abstract}{
We propose a new application of single molecule magnet crystals: their use as ``magnetic bubble chambers'' for the direct detection of sub-GeV dark matter. The spins in these macroscopic crystals effectively act as independent nano-scale magnets. When anti-aligned with an external magnetic field they form meta-stable states with a relaxation time that can be very long at sufficiently low temperatures. The Zeeman energy stored in this system can be released through localized heating, caused for example by the scattering or absorption of dark matter, resulting in a spin avalanche (or ``magnetic deflagration'') that amplifies the effects of the initial heat deposit, enabling detection. Much like the temperature and pressure in a conventional bubble chamber, the temperature and external magnetic field set the detection threshold for a single molecule magnet crystal. We discuss this detector concept for dark matter detection and propose ways to ameliorate backgrounds. If successfully developed, this detector concept can search for hidden photon dark matter in the meV - eV mass range with sensitivities exceeding current bounds by several orders of magnitude. 
}
\end{abstract}

\maketitle
\section{Introduction}
\label{sec:intro}

The continuing null results from experiments aiming to  directly detect dark matter (DM) with a mass above the GeV scale has stimulated theoretical and experimental excursions away from the WIMP paradigm. There exist well established experimental methods (ADMX, ADMX-HF) \cite{Ringwald:2012hr, Brubaker:2016ktl, Asztalos:2011bm} to probe DM particles such as axions and hidden photons with mass between 1 GHz - 100 GHz ($10^{-6} \text{ eV } - 10^{-4} \text{ eV}$).  New ideas employing nuclear magnetic resonance (CASPEr) \cite{Budker:2013hfa}, lumped element circuits (DM Radio) \cite{Chaudhuri:2014dla, Sikivie:2013laa, Kahn:2016aff}, and torsion balance/accelerometer \cite{Graham:2015ifn} (Eot-Wash) technologies have been proposed to search for ultra-light DM candidates in the mass range $10^{-22} \text{ eV } - 10^{-6} \text{ eV}$. 

While existing WIMP direct detection techniques have successfully lowered their thresholds, allowing for the detection of DM with mass greater than $\sim$100~MeV~\cite{Essig:2012yx,Guo:2013dt,Agnese:2016cpb, Chavarria:2014ika, Strauss:2016sxp}, it is theoretically possible that the DM is only somewhat lighter than conventional WIMPs and could lie in the mass range $10^{-3}$ eV - 100~MeV (for example, a dark photon, as considered in Section~\ref{sec:models}), a region for which no detection principle is known (recent proposals are referenced below). In such a mass region DM particles can deposit energies $\sim 10^{-3}$~eV - 10~eV through absorption (for bosonic dark matter with mass equal to {\it e.g.} a vibrational or electronic transition in a material) and inelastic scattering that are not large enough to be visible in conventional bolometric experiments.  On the other hand, protocols to search for ultra-light DM that do not rely on the deposited energy leverage the coherence of the ultra-light DM signal to build a measurable phase in an experiment. The coherence of the DM signal is inversely proportional to its mass and at masses greater than  $\sim 10^{-3}$ eV the coherence time is too small to employ phase accumulation techniques deployed to search for ultra-light DM. See Refs.~\cite{Essig:2011nj,Graham:2012su,Lee:2015qva,Essig:2015cda,Hochberg:2016sqx,Bloch:2016sjj,Hochberg:2016ntt,Derenzo:2016fse}  and Refs.~\cite{Hochberg:2015pha, Hochberg:2015fth, Hochberg:2016ajh, Schutz:2016tid, Knapen:2016cue} for recent proposals sensitive to DM energy depositions down to $\sim$\,eV and  $\sim10^{-3}$\,eV, respectively. 

Achieving sensitivity to such small energy deposits suggests the use of systems where some intrinsic energy gain is possible.  These systems, in general, require the storage of energy in a meta-stable state. The deposition of a small amount of energy can potentially lead to an avalanche in the system, where the initial energy deposit causes relaxation of the meta-stable state, releasing stored energy. This energy release can cause additional relaxation resulting in a runaway process that amplifies the initial deposited energy, enabling detection.   A well known system of this kind is the bubble chamber~\cite{PhysRev.87.665}, in which a liquid is maintained at a super-critical combination of temperature and pressure such that the energy deposited by an incident particle locally triggers the formation of a bubble. Because of the runaway reaction, the response of the detector is not proportional to the initial energy deposited. Thus, while these systems generally do not allow the measurement of the initial energy, very low thresholds, otherwise inaccessible, can be attained. Due to the low expected event rates, for a successful DM application it is important that the detector is stable for long ($\sim$\,months - year) timescales. It is also necessary to be able to reject radioactive backgrounds in the energy range of interest. 

In this paper we propose the use of a relatively newly discovered type of crystal ---single molecule (or molecular) magnets (SMMs)--- as a magnetic version of the bubble chamber that can be tuned to be sensitive to  $\sim 10^{-3}$ eV  - 10 eV energy deposits relevant for sub-GeV DM detection. This paper is organised as follows. In  Sec.~\ref{sec:overview}, we present a conceptual overview of the detector. Following this overview, in  Sec.~\ref{sec:smms} we review some relevant properties of SMMs, including details of their chemistry and synthesis. Sec.~\ref{sec:detector} outlines the proposed detector concept in greater detail. This includes the tuning of the parameters of such a `magnetic bubble chamber', the preparation of the crystal, and potential experimental backgrounds. To estimate the reach of such a detector, in Sec.~\ref{sec:models} we project sensitivity to a dark photon model, finding an improvement of  several orders of magnitude over existing stellar limits in the range $\sim 10^{-3}$ eV  - 10 eV. We conclude in Sec.~\ref{sec:conc}.   The present discussion is preliminary in nature and substantial work, both conceptual and experimental, will be required to better assess the feasibility.

\section{Overview}
\label{sec:overview}

SMMs are molecular crystals in which the molecules act as tiny, essentially non-interacting magnets~\cite{Lis:a19066,doi:10.1021/ja00015a057,cite-key}. Their study is currently a rapidly developing field of chemistry---many 100s of new SMMs have been created since their discovery in the 90s-- and, importantly, SMMs are easily and cheaply synthesized. Some SMMs are known to be fluorescent, opening the possibility that scintillating versions may be engineered, further enhancing their potential as particle detectors. 

\begin{figure}
\begin{center}
\includegraphics[width=12cm]{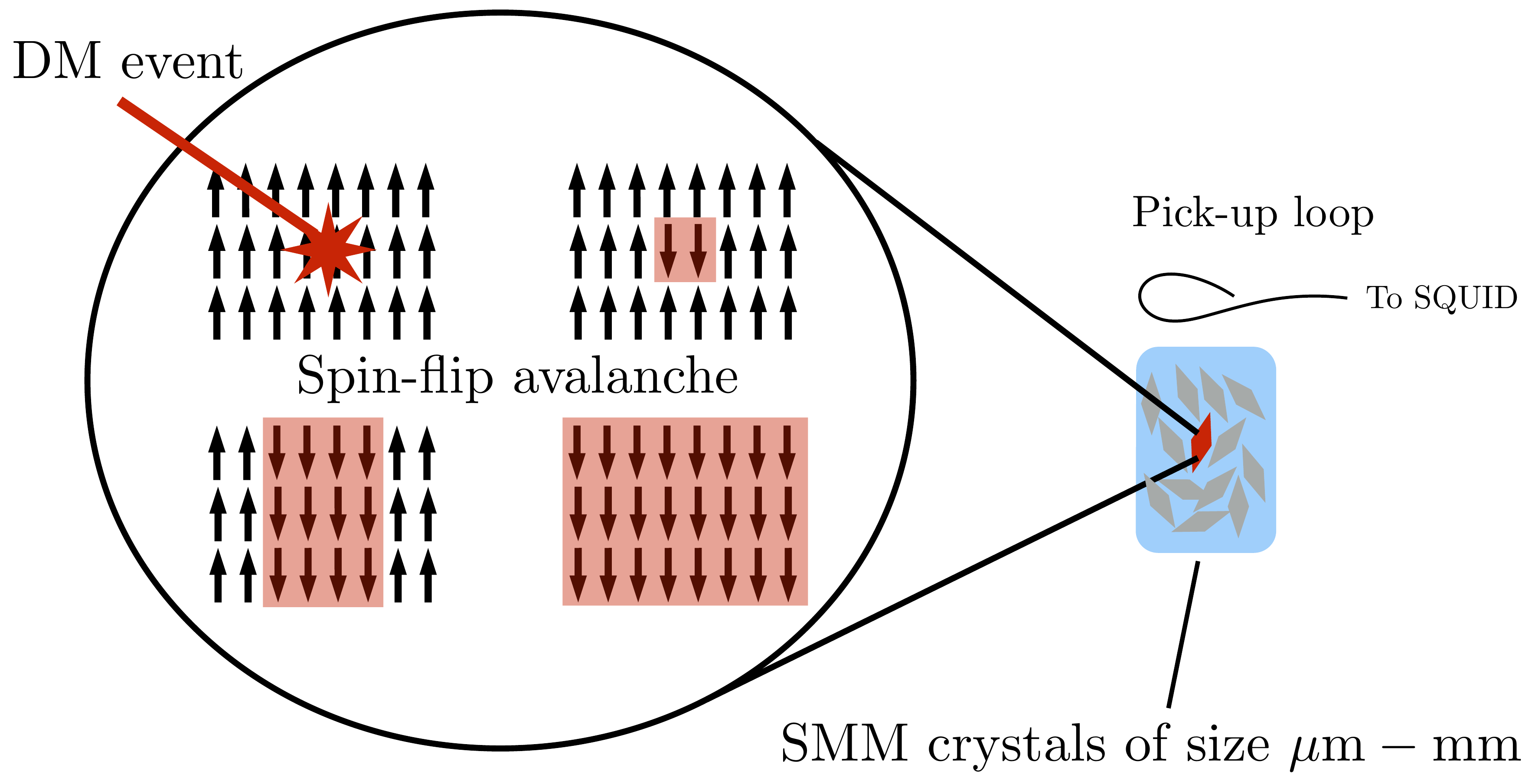}
\caption{DM detector concept based on magnetic deflagration in molecular nanomagnet crystals. A DM event that deposits energy in the form of heat ignites a spin-flip avalanche in the crystal which is detected by the change in magnetic flux through a pick-up loop.}
\label{fig:magdeflag}
\end{center}
\end{figure}
The basic idea for using SMMs as magnetic bubble chambers is as follows (see also Fig.~\ref{fig:magdeflag}). The crystal is prepared such that an O(1) fraction (in the simple preparation outlined here, 50\%) of the nano-magnets are anti-aligned with an external magnetic field, and so exist in a meta-stable state. This can be achieved simply by cooling the crystal and then applying an external magnetic field; at low temperatures ($\sim 0.1$\,K$-2$\,K) the magnetic relaxation (spin flip to the ground state) time can be of order years. We then exploit the key fact that the magnetic relaxation time is exponentially sensitive to temperature and the applied external field. Upon a deposit of energy, which can be of any form resulting in a local heating of the crystal ($\sim 10^{-3}$ eV  - 10 eV for the dark matter application considered here), the local magnetic relaxation time can drop below $\sim10^{-11}$\,s, which is comparable to the timescale of thermal diffusion in a few nm$^3$ sized region. The spins within this initial hot region will have time to relax, and the Zeeman energy released from this spin relaxation allows for the heating up of a larger region, causing other spins to flip. A runaway avalanche of spin flips ensues, a process known as ``magnetic deflagration'', first reported experimentally in Ref.~\cite{Paulsen1995}.  The tuning of the chemical molecular composition, temperature and magnitude of the external field such that the crystal sits in the right region for this process to occur upon a dark matter energy deposit, parallels the tuning of the pressure and temperature in a conventional bubble chamber.   

Unlike chemical burning, the magnetic deflagration is simply a spin wave which does not destroy the sample; yet a mechanism to ``quench'' the deflagration is required, so that a single event, whether due to signal or background, does not extend to the entire detector, resulting in an unacceptable dead-time.  Two quenching mechanisms can be considered: a) in a large SMM crystal, the magnetization can be monitored with precision magnetometry and, once a spin avalanche signal is detected, the magnetic field providing the energy for the deflagration can be rapidly turned off, effectively quenching the cascade process, in analogy to recompression in a bubble chamber; b) the material can be prepared in small sized grains with sufficiently poor thermal coupling between them, such that deflagrations are limited to single grains.  Which technique will prove more effective will have to be decided after more work on the material synthesis and detector implementation are carried out.   For the purpose of this paper we will simply show that precision magnetometers~\cite{Asztalos:2011bm} can observe the magnetic signal due to spin reversal in a $\sim 10^3\;\rm \mu m^3$ region in $\sim \rm 10^{-7} \;s$  and that the magnetic field can be turned off on a similar time-scale.  This provides a very substantial tolerance to dead-times due to background events. 

In analogy with a conventional bubble chamber, the quenching procedure also provides a reset mechanism: at zero magnetic field, the raised temperature (from Zeeman energy) in the bubble region will serve to re-equilibrate the spins to an equal split  between potential wells. Upon waiting further sufficient time such that the heat dissipates---about $10^{-5}$\,s---the magnetic field can be turned on again and the detector is reset.

Finally, a larger fraction of anti-aligned nano-magnets could potentially be achieved by cooling while applying the magnetic field such that all spins are aligned, and then quickly reversing the field direction. For simplicity we do not consider this possibility here: {\it i)} in any case, a large fraction of spins could reverse orientation at the point of zero magnetic field during the reversal, and {\it ii)} any resulting internal magnetic field of the crystal would have to be precisely cancelled in the quenching mechanism.

\section{Single Molecule Magnets}
\label{sec:smms}

An introduction to SMMs that includes a review of magnetic deflagration can be found in {\it e.g.} \cite{2010ARCMP...1..109F}. Here we provide details that are necessary for our purposes, and introduce three parameters which play a role in the tuning of the proposed device: the effective magnet spin, $J$, the energy barrier, $U$, and the time constant relative to the fastest possible relaxation, $\tau_0$. 

A molecule in an SMM crystal typically consists of a magnetic core surrounded by a non-magnetic shell. The magnetic core may consist of multiple metal ions coupled through non-magnetic centers ($e.g.$ oxygen dianions), or the core may be a single metal ion. This core has large effective spin $J$ which, in case of multiple metal ions, arises from strong (super)exchange interactions coupling the smaller spins of individual ions, or in the case of a single metal ion, arises from the coupling of spin and orbital angular momentum~\cite{kahn1993molecular,Gatteschi:993655}. The magnetic core is surrounded by a non-magnetic shell (typically organic ligands such as acetate) which acts to separate the magnetic cores from each other such that the exchange interactions between the core spins are anomalously weak. This architecture permits the possibility of magnetic deflagration---the negligible intermolecular magnetic interactions allows for the spins to release locally stored Zeeman energy. On the other hand, the non-magnetic shell provides the thermal contact necessary to transport the released heat from one molecule to the next, encouraging magnetic relaxation of neighboring molecules. 

\begin{figure}
\begin{center}
\includegraphics[width=16cm]{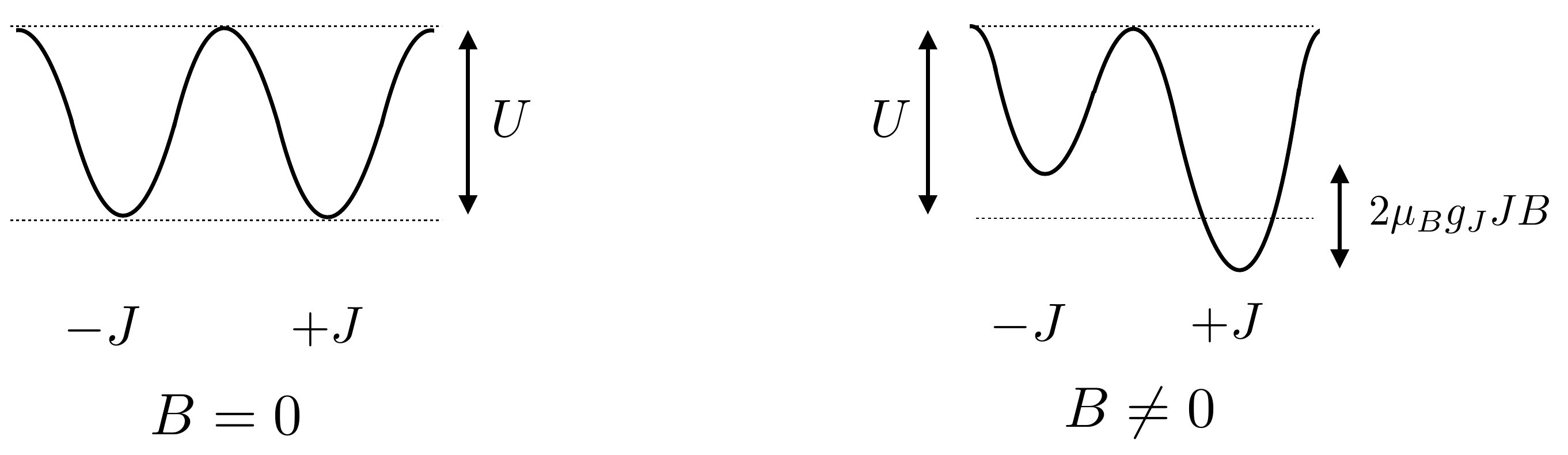}
\caption{{\it Left:} Potential felt by individual molecular magnets in the crystal. {\it Right:} Lifting of degenerate ground states in an external magnetic field.}
\label{fig:potwell}
\end{center}
\end{figure}

The effective spin $J$ contains (2$J$+1) $M_{J}$ states, where the value of $M_{J}$ denotes the projection of $J$ along a preferred axis. Spin-orbit coupling and the interaction between the effective spin $J$ and the crystal field causes a splitting of these $M_{J}$ states such that an energy barrier, $U$, arises between $M_{J}$ = $\pm J$ and $M_{J}$ = 0 (for integer $J$) or 1/2 (for half-integer $J$), with $M_{J}$ = $J-1$, $J-2$... states at predictable energies in between (see \cite{2010ARCMP...1..109F} for an effective Hamiltonian that describes this splitting). This manifold of $M_{J}$ states is typically depicted as a double-well potential (figure \ref{fig:potwell}). In the absence of a magnetic field, the $M_{J} = -J$ and $M_{J} = +J$ states are degenerate.  An applied magnetic field, $B$, stabilizes one side of the well and destablizes the other. The difference in energy between $M_{J} = -J$ and $M_{J} = +J$ states (the Zeeman energy) is given by $2 \mu_B g_{J} J B$, where $g_{J}$ is the Land\'e g-factor of the molecule. A molecule in the newly created metastable state is protected from instant decay if $kT \ll U$; when it does decay it releases a phonon equal to the Zeeman energy.

The relaxation time $\tau$ for spins to flip in thermal equilibrium is given by
\be
 \frac{1}{\tau} =\frac{A_1}{1+A_2 B^2} + C B^2T+DT^n+\frac{1}{\tau_0}\exp(-\widetilde{U}(B)/kT) \,,
\label{eq:taudep}
\ee
where the four terms describe relaxation by quantum tunneling~\cite{ANIE:ANIE200390099}, direct, Raman, and Orbach~\cite{0370-1328-77-2-305, Orbach458, Abragam:1625648} relaxation, respectively. The constants, $A_1$, $A_2$, $C$, $D$, and $\tau_0$ are unique to each SMM;  $B$ is the applied field; $\widetilde{U}(B)$ is the energy barrier described previously, modified by the $B$ field. To first approximation,
\be
\widetilde{U}(B)=U-\frac{1}{2} \Delta E_{\text{Zee}} \,,
\label{eq:zeesplit1}
\ee 
where we defined the Zeeman splitting,
\be
\Delta E_{\text{Zee}} = 2 \mu_B g_{J} J B \,.
\label{eq:zeesplit2}
\ee
The four terms are listed in the order in which they are the dominant relaxation process going from low to high temperature. In the absence of an applied field, relaxation time at low temperature is determined by relaxation through a quantum tunneling mechanism. The tunneling relaxation mechanism involves direct interactions between states on either side of the double well potential and relaxation occurs without the release of a phonon. The strong field dependence~\cite{PhysRevLett.80.612} removes tunneling as a viable relaxation pathway in even a modest field. The remaining processes are phonon dependent~\cite{0370-1328-77-2-305, Orbach458, Abragam:1625648}. The direct process emits a single phonon equal to the Zeeman energy as the molecule relaxes from $|+J\rangle$ to $|-J\rangle$ and is typically only appreciable in SMMs with single metal centers in large applied fields. Raman and Orbach relaxation processes are both two-phonon processes, where the absorption and emission of phonons (corresponding to moving up and down the ladder of $M_J$ states) proceeds through either virtual (Raman) or real (Orbach) excited states. The power dependence of Raman relaxation has been described for a number of systems, and is generally large ($n = 5-9$) due to the increasing availability of phonons for relaxation at higher temperatures~\cite{PSSB:PSSB2221170202}.

\begin{figure}
\begin{center}
\includegraphics[width=8cm]{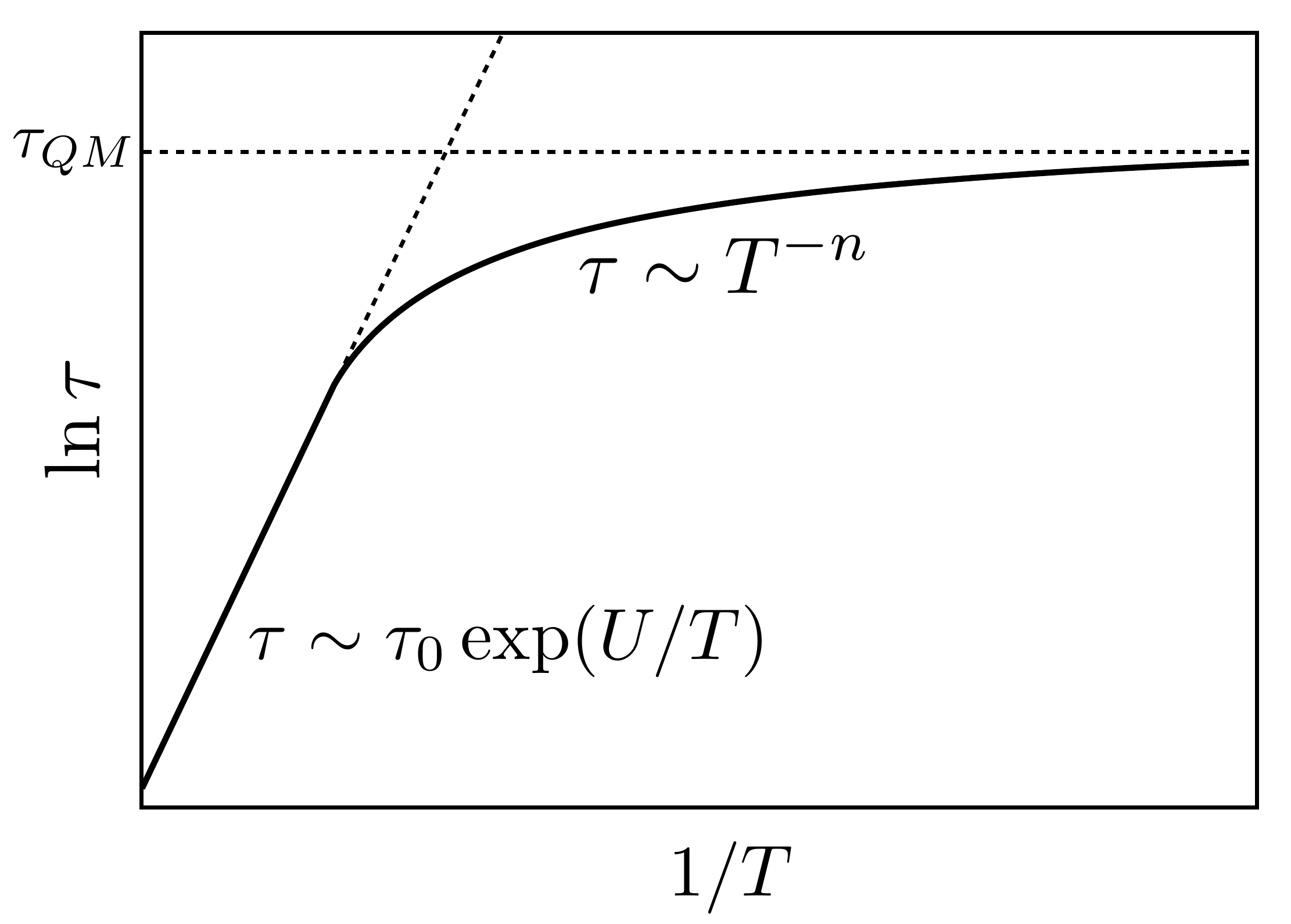}
\caption{Typical behaviour of magnetic relaxation time as a function of temperature for an SMM. The temperature dependence that dominates at high and intermediate temperatures is indicated. The low temperature behaviour approaches the limiting quantum mechanical tunneling relaxation time, denoted $\tau_{QM}$.} 
\label{fig:tempdep}
\end{center}
\end{figure}

The relaxation time is a strong function of both temperature and the applied magnetic field. At low temperatures ($\sim$ 0.1 K)  and high magnetic fields ($\gtrsim$ 0.1 T), the relaxation times are very long ($\sim$ months - years) while at higher temperatures ($\sim$ 30 K) the relaxation time can be as short as   $\sim \tau_0 \sim 10^{-10} -10^{-14} \, s$.  This behavior implies that localized heating  can lead to a rapid rise in spin relaxation, releasing stored Zeeman energy. The released energy further heats up the  sample, resulting in additional spin relaxation, resulting in the avalanche that is triggered by a localized energy deposit (for example, by DM scattering or absorption).

The dominant relaxation mechanism depends both on the type of SMM and the temperature. Consider, for example, Mn$_{12}$-acetate, a SMM with a magnetic core of 12 manganese ions. The primary relaxation pathways for Mn$_{12}$-acetate are~\cite{ANIE:ANIE200390099}: tunneling (ground state to ground state), thermally assisted tunneling (excitation partially up the double well and then tunneling from this excited state), or Orbach relaxation (moving from one state to the next up the double well through successive absorptions of phonons). Relaxation at low temperature is only through tunneling and is therefore very slow (on the order of months). At higher temperatures Orbach relaxation dominates and its speed is only limited by the pre-exponential term $\tau_0$, which can be very short. Thus, a localized change in temperature can rapidly change the relaxation time, permitting an avalanche.

The majority of SMMs can be chemically categorized in the following way: SMMs with a single metal ion versus multiple metal ions, and those metal ions can be d-block metals (transition metals) or f-block metals (Lanthanides/Actinides). The parameters $J$, $U$ and $\tau_0$ broadly characterize the properties of a given SMM, and in the 100s of SMMs synthesized in the past two decades their values vary widely: 
\be
\text{1} <&\,J\,&<\, 45 \,,\nonumber \\
\sim 1K <&\,U\,&<\, 1815  K \,,\nonumber \\
\sim 10^{-6}s < &\,\tau_0\,& <\sim 10^{-14}s\,. \nonumber 
\ee 
As we will see in section  \ref{sec:detector}, these quantities  set the  threshold energy necessary to trigger magnetic deflagration.  Some properties of the better known molecules in these categories are given in Tab.~\ref{tab:smms}. Our initial focus is on SMMs featuring multiple transition metals, and a brief description of several of these may illustrate the ease of access and tunability of these materials.

\begin{table}
\centering
\begin{tabular}{|l|c|c|c|c|}
\hline
SMM & $J$  & $\tau_0$\,[s] & $U$\,[K]  &  Ref.\\
\hline

{\it Mononuclear Transition Metal SMMs }&&&&\\
$[$K(2.2.2-crypt)$][${\bf Fe}(C(SiMe$_3$)$_3$)$_2$$]$& 7/2 & $4.5\times10^{-10}$  & 354 & \cite{doi:10.1021/ic402013n}\\
\hline
{\it Multinuclear Transition Metal SMMs}&&&&\\
 {\bf Mn}$_{12}$O$_{12}$(O$_2$CCH$_3$)$_{16}$(H$_2$O)$_4$$\cdot$HO$_2$CCH$_3$$\cdot$4H$_2$O& 10 & 2.1$\times10^{-7}$  & 61 & \cite{Sessoli1993}\\
{\bf Mn}$_{12}$O$_{12}$(O$_2$CC$_6$H$_4$-$p$-Me)$_{16}$(H$_2$O)$_4$$\cdot$HO$_2$CC$_6$H$_4$-$p$-Me& 10 & $2.0\times10^{-10} $ & 38 & \cite{doi:10.1021/ic000911+}\\
{\bf Mn}$_{12}$O$_{12}$(O$_2$CC$_6$H$_4$-$p$-Me)$_{16}$(H$_2$O)$_4$$\cdot$3H$_2$O & 10 & $7.7\times10^{-9}$ & 64 & \cite{doi:10.1021/ic000911+}\\
{\bf Mn}$_6$O$_2$(sao)$_6$(O$_2$CPh)$_2$)$_2$(MeCN)$_2$(H$_2$O)$_2$ & 4 &$6.6\times10^{-8}$  & 24 & \cite{B822235E}\\
{\bf Mn}$_6$O$_2$(Et-sao)$_6$(O$_2$CC(CH$_3$)$_3$)$_2$(EtOH)$_5$ & 6 & $3.0\times10^{-8}$  & 30 & \cite{doi:10.1021/ja070411g}\\
{\bf Mn}$_6$O$_2$(Et-sao)$_6$(O$_2$CC$_6$H$_4$(CH$_3$)$_2$)$_2$(EtOH)$_6$ & 12 & $2.0\times10^{-10}$  & 86 & \cite{doi:10.1021/ja068961m}\\
{\bf Fe}$_4$(CH$_3$C(CH$_2$O)$_3$)$_2$(dpm)$_6$ & 5 & $2.1\times10^{-8}$  & 17 & \cite{doi:10.1021/ja0576381}\\\hline
{\it Mononuclear Lanthanide SMMs} &&&& \\
$[${\bf Dy}(O$^\text{t}$Bu)$_2$(C$_5$H$_5$N)$_5$][BPh$_4$]  &15/2 & $1.2\times10^{-12}$ & 1815 & \cite{ANIE:ANIE201609685}\\
\hline {\it Multinuclear Lanthanide SMMs} &&&&\\
$[$K(18-crown-6)$][$$\{$$[$(Me$_3$Si)$_2$N$]$$_2$(THF){\bf Dy}$\}$$_2$($\mu$-$\eta^2$:$\eta^2$-N$_2$)$]$ & 29/2 & $8.0\times10^{-9}$  & 178 & \cite{Rinehart2011}\\
$[$K(18-crown-6)$][$$\{$$[$(Me$_3$Si)$_2$N$]$$_2$(THF){\bf Tb}$\}$$_2$($\mu$-$\eta^2$:$\eta^2$-N$_2$)$]$ & 23/2 & $8.2\times10^{-9}$  & 327 & \cite{doi:10.1021/ja206286h}\\

\hline
\end{tabular}
\caption{Properties of SMMs from four different families. The bolded terms are the metal ions that comprise the magnetic core. The non-bolded terms describe chemical groups in the non-magnetic shell that can be substituted with similar chemical groups to tune the SMM properties.}
\label{tab:smms}
\end{table}

\begin{figure}
\begin{center}
\includegraphics[width=7.3cm]{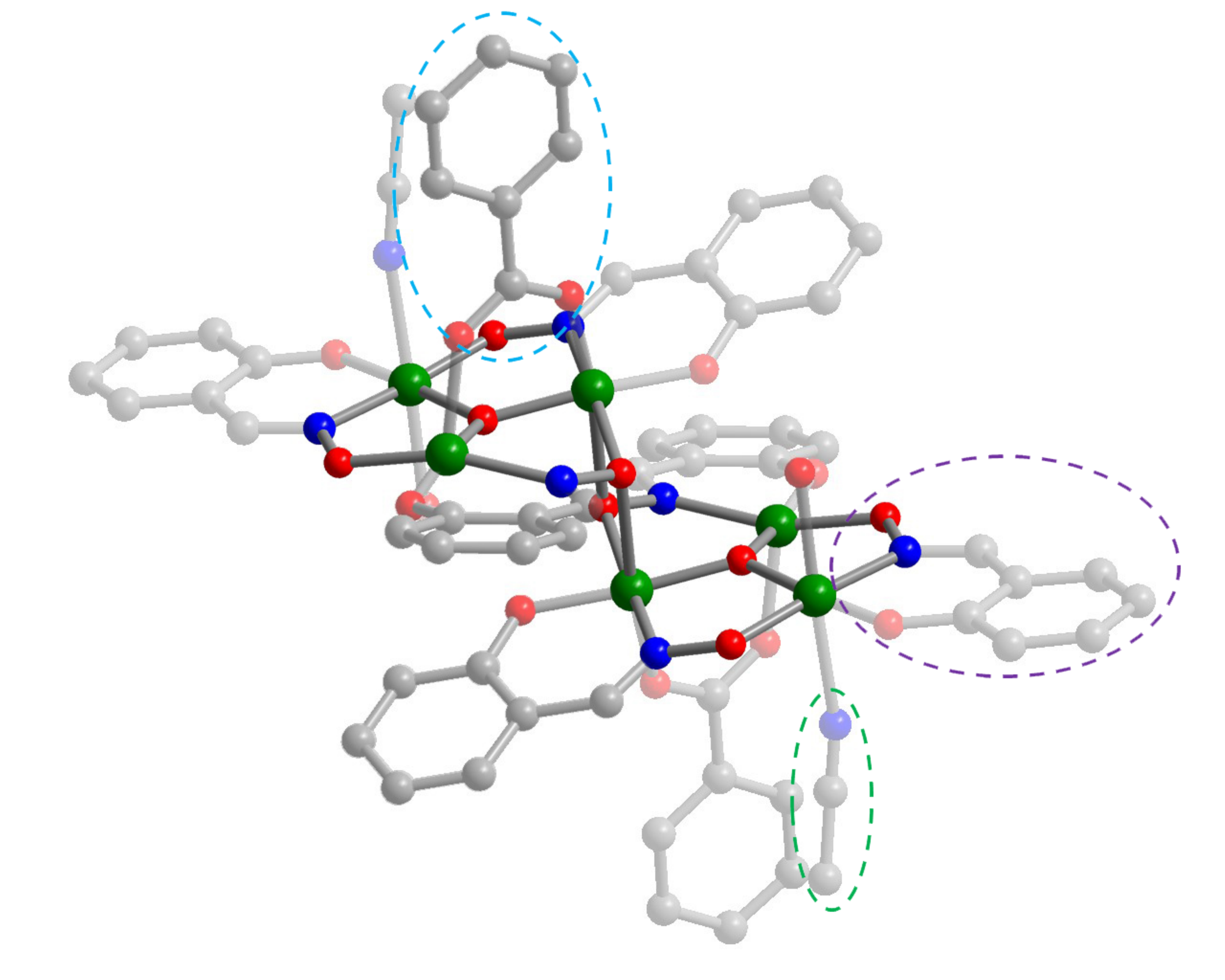}
\qquad\qquad
\includegraphics[width=7.3cm]{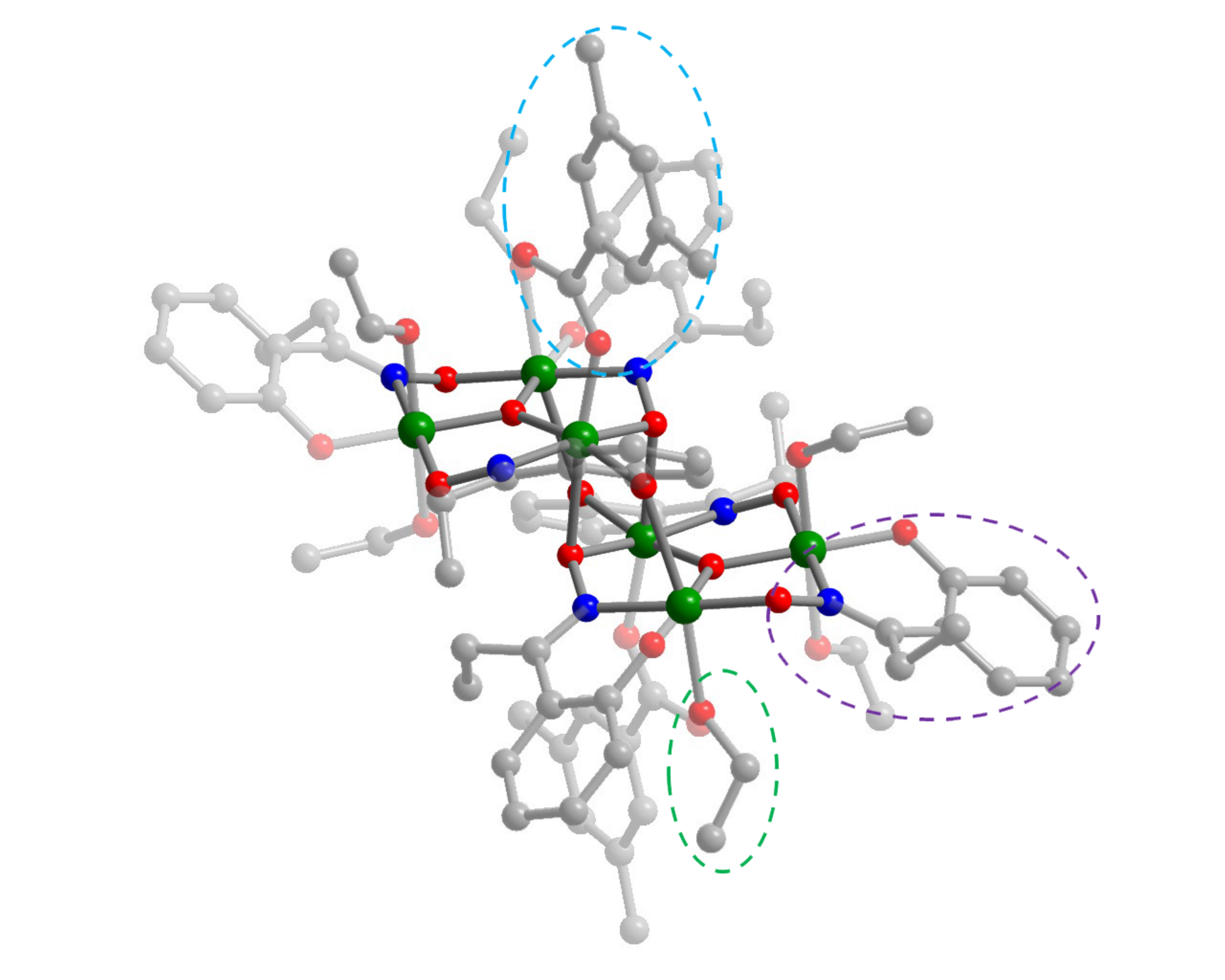}
\caption{Molecular structures for [Mn$_6$O$_2$(H-sao)$_6$(O$_2$CPh)$_2$(MeCN)$_2$(H$_2$O)$_2$] (left) and [Mn$_6$O$_2$(Et-sao)$_6$(O$_2$CPh(CH$_3$)$_2$)$_2$(EtOH)$_6$] (right). Green, red, blue, and grey spheres represent manganese, oxygen, nitrogen, and carbon, respectively. Hydrogen atoms have been omitted for clarity. In each structure the Mn$_6$ unit has been highlighted. The tunable parts of the molecules have been circled with H-sao and Et-sao in purple circles, O$_2$CPh and O$_2$CPh(CH$_3$)$_2$ in blue circles, and MeCN and EtOH in green circles.}
\label{fig:chemmole}
\end{center}
\end{figure}

We highlight the chemical control over the properties of SMMs through an example within the Mn$_6$ family of SMMs---one of the best studied families of single-molecule magnets~\cite{B822235E}. These molecules feature two trigonal [Mn$_3$O]$^{7+}$ cores each with $S=2$ or $S=6$ which couple to each other such that the spin of the Mn$_6$ SMM can range from $J=4$ to $J=12$. The general synthetic strategy is as follows: Mn(ClO$_4$)$_2$$\cdot$6H$_2$O is dissolved in methanol (MeOH), ethanol (EtOH), or acetonitrile (MeCN). To this solution is added a molar equivalent of of salicylaldoxime (where salicylaldoxime is abbreviated H-sao, and modified versions of this ligand are Me-sao or Et-sao when methyl (Me) or ethyl (Et) groups replace a proton (H) in a particular position), a carboxylic acid (RCO$_2$H where R can be a variety of organic groups), and a base (e.g. sodium methoxide). From this approach one can make, for example, [Mn$_6$O$_2$(H-sao)$_6$(O$_2$CPh)$_2$(MeCN)$_2$(H$_2$O)$_2$]~\cite{B822235E} or [Mn$_6$O$_2$(Et-sao)$_6$(O$_2$CPh(CH$_3$)$_2$)$_2$(EtOH)$_6$]~\cite{doi:10.1021/ja068961m}. While both contain very similar Mn$_6$ cores, the change in organic groups changes bond angles sufficiently such that the magnetic properties change substantially; the former molecule has $J=4$, $\tau_0=1.7\times10^{-8}$\,s, and $U=34$\,K while the latter has $J=12$, $\tau_0=2\times10^{-10}$\,s, and $U=86$\,K.  See figure~\ref{fig:chemmole} for the molecular structure of these compounds. The materials are relatively inexpensive. By simply scaling up the known synthesis of [Mn$_6$O$_2$(Et-sao)$_6$(O$_2$CPh(CH$_3$)$_2$)$_2$(EtOH)$_6$], a kilogram of material could be made for less than 3000 US dollars.

\section{Detector concept and requirements}
\label{sec:detector}

As outlined in section~\ref{sec:overview}, a viable detector for sub-GeV-mass DM particles has to achieve a sufficient sensitivity with negligible background.  We can gain sensitivity to DM energy deposits in the region $\sim 10^{-3}$ eV  - 10 eV  through requiring that this be the threshold for triggering a deflagration.

The magnitude and properties of the background produced by various types of radiation in the very low energy regime of interest here is mostly unknown.   It can be generally argued (e.g.~\cite{2012JLTP..167.1081P}) that the event density from a $\sim$ flat Compton background in an energy region of very small extent, such as the one considered here, is exceedingly small. Of course, it is possible that minute energy depositions from processes other than Compton scattering may become important in the extremely low-energy regime of interest here.  This issue is common to all possible concepts for detectors attacking an entirely unexplored energy region and only experimental tests with actual materials will reveal which background sources may be problematic and how to reduce them to a manageable level.  

The detector proposed here integrates signals above a threshold and hence does not provide a proper energy measurement. From the point of view of background rejection, this limitation can be mitigated by nothing that in most cases it is expected that energy depositions larger than those of interest will leave a ``trail'' of deflagrations.  This is particularly true if the detector is finely segmented in grains (case b) in section~\ref{sec:overview}), allowing a high-energy veto to be applied.  The possibility of adding scintillation to the techniques used to veto higher-energy events has also been already mentioned.  

An important issue is that the detector should be `live' for a significant fraction of the time. The reset following a background event can in principle occur within $\sim 10^{-5}$\,s (as outlined in the overview above). Radon backgrounds in existing experiments are at a rate much lower than this, {\it e.g.} in LUX at a rate of $\sim$1/m$^2$/s which corresponds to $\sim$1 background radon event per 100s in a kg SMM detector; thus, at least for known backgrounds, the possibility of significant live time seems achievable.

It is also important that the system is otherwise stable, i.e. a negligible rate of deflagrations occur without an interaction in the detector.   This is the equivalent of the ``dark rate'' in a photodetector.   While, given the large spin density, there will always be some non-negligible probability for a single spin to flip in a SMM crystal, parameters can be chosen so that a single spin flip cannot set off an avalanche, viz. the size of the initial hot region (i.e. our thresholds) needs to be at least a few spins (few angstroms) ``long". In this way, the probability for all these spins to simultaneously flip is very low, given the long relaxation time at low temperature, and the dark rate can be arbitrarily reduced.  On the other hand, the DM scattering deposits more energy than the Zeeman energy released by a single spin flip, raising the temperature of a larger region, triggering a deflagration. 

We now conceptually discuss, in turn, the detector preparation, threshold tuning, deflagration quenching and reset.

\subsection{Detector preparation}
Preparation for an operating run consists of cooling the system to temperatures $\sim0.1$~K, and then applying a magnetic field ($\sim 0.1-1$~T) to the sample. After this procedure, half of the spins will exist in the meta-stable well of the potential, as a result of the very long magnetic relaxation time at low temperature. 

We require there to be no significant loss of usable detector after the preparation stage ({\it i.e.} that roughly half the spins end up in the meta-stable state---in particular that no avalanche occurs). Magnetic relaxation times under an applied magnetic field are typically longer than in the zero field case. This is due to a mismatch of energy levels of the two potential wells, thus reducing the probability of quantum mechanical tunneling~\cite{PhysRevLett.76.3830}. The longest relaxation times experimentally determined are of order years under applied magnetic fields $\sim1$\,T, see e.g.~\cite{doi:10.1021/jacs.5b13584}. However, for e.g. the same SMM in~\cite{doi:10.1021/jacs.5b13584}, the low temperature, zero field relaxation time is $\sim10$\,s. Assuming that the time taken to ramp up the magnetic field is $\sim10^{-5}$\,s, then, even if this relaxation time remained at the zero field value (which it will not, because it has a strong dependence on the magnetic field), there is a $\sim10^{-6}$\, probability for a single spin to flip. This is still below what could cause an avalanche, since the probability for a region of radius a few spins long to all flip is still negligible in a crystal with $10^{15}-10^{18}$ molecular magnets. In reality, we expect that a negligible fraction of spins will flip in the ramp-up, such that the fraction of spins in the meta-stable state remains $\sim 0.5$, with no avalanche occurring during preparation.

\subsection{Threshold tuning}
We consider DM depositing heat in the crystal; this could be via a scattering or an absorption mechanism. The heat causes a local increase in temperature, and we exploit the fact that the relaxation times of SMMs are exponentially sensitive to temperature. If this high temperature relaxation time is short compared to the thermal diffusion time scale from that region, then this will cause the spins in that region to relax from their metastable state to the stable state, releasing their stored Zeeman energy. 

Consider an energy deposit, $E_0$, in a region of size $R^3$. The temperature of this region is raised by $\Delta T$,
\be
\frac{dE}{dT} = C(T) \sim c_0 R^3 T^3 \implies \Delta T = \bigg(\frac{E_0}{c_0 R^3} \bigg)^{1/4} \,,
\label{eq:dedt}
\ee
where $c_0$ is the volume-specific heat capacity, and where we take a typical temperature dependence for the heat capacity of a crystal at low temperature (this Debye behaviour has been observed in SMMs down to temperatures of 1\,K see e.g.~\cite{B603738K, doi:10.1021/ic010567w}). 
The magnetic relaxation time of the spins in this hot region becomes
\be
\tau\simeq \tau_0\exp\left(\frac{U- \frac{1}{2}\Delta E_{\text{Zee}}}{ \Delta T}\right)\,,
\ee
where the energy barrier appearing in the Arrhenius law part of eq.~\eqref{eq:taudep} is modified by the Zeeman splitting, as per eqs.~\eqref{eq:zeesplit1},~\eqref{eq:zeesplit2}. Typical values of $E_{\text{Zee}}$ range from $0.01-0.1$\,meV.

If the initial hot region is of radius $R$, the thermal diffusion time scales $\propto R^2$, while the spin relaxation time is independent of $R$. Thus, for a sufficiently large $R$ ({\it i.e.} a region sufficiently ``long'' in spins), the spins will relax before the heat dissipates, in analogy to the critical radius for bubbles to form in a conventional bubble chamber. So for a given choice of parameters, there is a critical size $R$ that needs to get heated up for the spins to relax. The energy released from this spin relaxation will then cause other spins to flip, heating up a larger region. The heat from this larger region will take longer to dissipate, and thus all the spins there will also relax, resulting in the deflagration.

The temperature rise $\Delta T$ caused by this interaction must reduce the magnetic relaxation time $\tau\lesssim \tau_{D}$ where the thermal diffusion time $\tau_{D} \sim  R^2/\alpha$, $\alpha$ being the thermal diffusivity. Taking $\alpha \sim 10^{-7}\,$m$^2$/s at 1K (typical of a number of materials, including water and at least one SMM \cite{thesis}), we have for $R \sim 3$\,nm, $\tau_{D}\sim 10^{-10}$s. In such a material, if an energy deposition within a region of radius $3$\,nm increases the temperature of that region to cause spin relaxation within $\sim10^{-10}$\,s, the spins can flip and so this energy deposit could trigger a deflagration wave.

\begin{figure}
\begin{center}
\includegraphics[width=10cm]{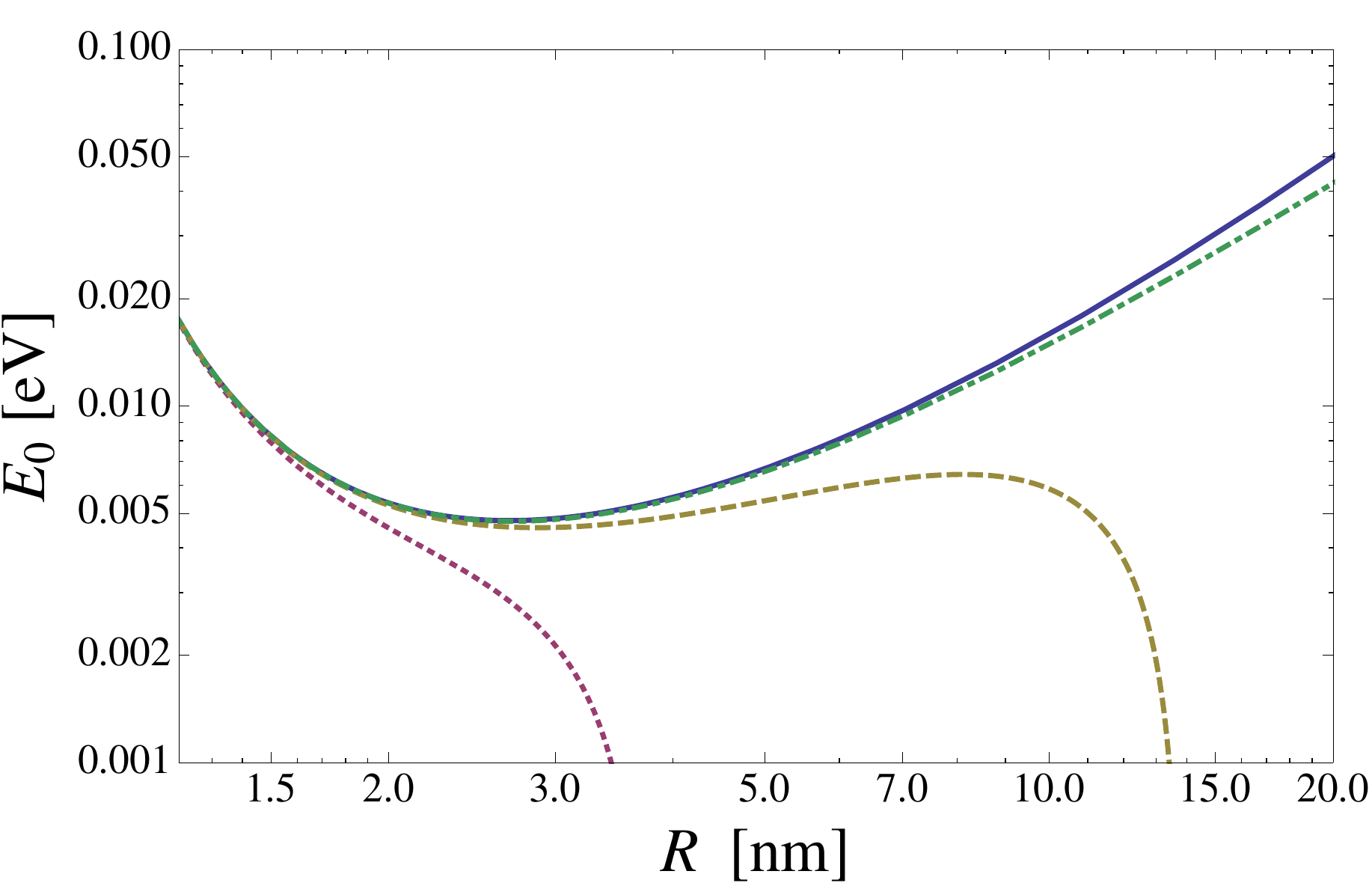}
\caption{Tuning curves for a SMM with $U=50$\,K ($\simeq4.3$\,meV), $\tau_0=5\times10^{-12}$\,s, and for $\Delta E_{Zee}=(0,0.001,0.01,0.1)$\,meV (blue solid, green dot-dash, yellow dash, red dot).}
\label{fig:tuning}
\end{center}
\end{figure}

Under what conditions does the spin avalanche occur? To address this we turn to aspects of tuning the detector {\it i.e.} the balance between the parameters $U$, $J$, $\tau_0$ {\it etc.}. The condition for spins to flip in a region of size $R$ is $\tau\lesssim \tau_D$, that is,
\be
\tau_0 \exp\left(\frac{U-\frac{1}{2}\Delta E_{\text{Zee}}}{\Delta T}\right) &\lesssim&\frac{R^2}{\alpha} \,.
\ee
Substituting for $\Delta T$ using eq.~\eqref{eq:dedt}, this implies that the energy deposit in this region must satisfy
\be
E_0 \gtrsim \frac{c_0 R^3 (U-\frac{1}{2}\Delta E_{\text{Zee}})^4}{\ln\left[\frac{R^2}{\tau_0\alpha}\right]^4}\,.
\label{eq:ethresh}
\ee
The solid blue curve in Fig.~\ref{fig:tuning} illustrates this inequality for an SMM with parameters $U=50$\,K ($\simeq4.3$\,meV), $\tau_0=5\times10^{-12}$\,s,  and $\Delta E_{\text{Zee}}=0$ (zero field case); we take typical values for the parameters $\alpha=10^{-7}$\,m$^2$/s, and $c_0=2\times 10^{-9}$\,eV/K$^4$/nm$^3$. We see the presence of an energy gap $\sim U^4 (\tau_0 \alpha)^{3/2} c_0$; the threshold would be higher/lower if we pick a SMM whose with larger/smaller values of $U$ or $\tau_0$. 

We can gain a conservative estimate how much Zeeman energy is needed to fuel an avalanche by asking what $E_0$ would be required to satisfy $\tau\lesssim \tau_D$ if we allow that the Zeeman energy inside the region has already been released and has all converted to heat in this region; if we find that progressively less input energy is needed, we find that an avalanche will occur. We set $E_0\to E_0+\rho_s R^3 \Delta E_{\text{Zee}}$, where $\rho_s$ is the density of nanomagnets ($\sim 1/\text{nm}^3$). This modifies eq.~\eqref{eq:ethresh},
\be
E_0 &\gtrsim&\frac{c_0 R^3 (U-\frac{1}{2}\Delta E_{\text{Zee}})^4}{\ln\left[\frac{R^2}{\tau_0\alpha}\right]^4} - \rho_s R^3 \Delta E_{\text{Zee}} \,,
\ee
which is plotted as the broken lines in Fig.~\ref{fig:tuning}, for different values of $\Delta E_{\text{Zee}}$. The Zeeman energy pulls the `tuning' curve over, eventually to negative values of $E_0$, indicating the onset of the avalanche. 

Hence, if we take a crystal with $\tau_{0} \sim 10^{-11}$\,s, with $U \sim 50$\,K, we can see that an energy deposit of $\sim 0.01$ eV within $\sim 3$\,nm would be sufficient to flip the spins in this region. This will trigger an avalanche in the presence of an external magnetic field $\sim 0.1 - 1$\,T., {\it i.e.} such that $\Delta E_{\text{Zee}}=0.01-0.1$\,meV for nanomagnets of spin $J$ in the range $10-50$. 

There are a number of magnets with parameters in this desired range, though the experimental reports do not in general contain all the necessary data ({\it i.e.} such as the actual thermal diffusivity of the material, or zero and non-zero magnetic field relaxation times) necessary to identify a particular material that can be used for dark matter detection. The use of a higher threshold material (either through the chemical synthesis or through adjusting the magnitude of the external field) would be desirable to probe the higher end of the energy region we address.

\subsection{Deflagration quenching}

Under the assumption of a large SMM crystal (case a) in~\ref{sec:overview}) we now examine in some detail the quenching process.   When the external magnetic field is turned off, the Zeeman energy release that drives the deflagration is removed, quenching the process. The speed of the deflagration wave is  $\sim 1\,{\rm nm}/\tau$ where $\tau \sim \tau_0$ is the actual relaxation time of the spins, 
with the ``burning'' occurring layer by layer. This yields a deflagration front moving with a sub-sonic speed $\sim 100$ m/s through the material. It can be seen that with a magnetometer of sensitivity  $\sim 0.1 \text{ fT}/\sqrt{\text{Hz}}$ \cite{Asztalos:2011bm}, the spin-flips caused by this front would be visible after a period of $\sim 10^{-7}$\,s, once it burns through $\sim 10 \, \mu \text{m}$ of the material.  After this time, if the magnetic field is shut off, the flame front will turn off, protecting the remainder of the material.

When switching off the magnetic field, the energy stored in it needs to be removed. If the device is operated with a magnetic field $\sim 0.1$~T and a volume of $\sim \left(10 \text{ cm}\right)^3$ (corresponding to a $\sim$~kg scale target mass), the energy stored in the field is $\sim 10$~J that can be dissipated in a suitable resistor in $10^{-7}$\,s.

\subsection{Reset} As noted above, the turnoff of the magnetic field is a mechanism which naturally resets the detector.  This may occur periodically to quench deflagrations in relatively large crystals, or may become periodically necessary when a certain number of grains have changed state and have become inactive.

\subsection{Calibration} The experimental calibration of the energy threshold for the onset of spin avalanches in the detector could be performed in a way similar to existing experimental investigations of magnetic deflagration in SMMs---see Ref.~\cite{2013arXiv1302.5100S} for a review of these techniques---where {\it e.g.} a heater, surface acoustic waves or a current pulse are used to provide controlled, temperature-driven avalanches.

\section{Absorption of a dark vector}
\label{sec:models}

To illustrate the potential capability of such a detector, we consider the direct absorption of a dark vector particle, which is a well-motivated (simplified) model of dark matter, involving one of the few possible renormalizable  DM-SM interaction terms (see also {\it e.g.} Ref.~\cite{Graham:2015rva} for a viable cosmological production mechanism such that vector particles in the mass range considered here can constitute cold dark matter). We estimate the sensitivity for absorption of a dark vector in the meV--eV mass range, where the hidden photon mass equals that of a vibrational resonance of the molecules in the crystal, or a low-energy electronic transition in the molecule. We are specifically targeting energies below that of ionization, where we expect existing technologies to be competitive.

The Lagrangian for a dark vector with Stuckelberg mass $m_V$ is 
\be
\mathcal{L}= -\frac{1}{4} F_{\mu\nu}^2 - \frac{1}{4} V^{2}_{\mu\nu} - \frac{\kappa}{2} F_{\mu\nu} V^{\mu\nu} + \frac{m_V^2}{2}V_\mu V^\mu + e J_{\text{em}}^\mu A_\mu\,, 
\label{eq:lagrang}
\ee
where $F_{\mu\nu}$ and $V_{\mu\nu}$ are field strength tensors for the photon, $A_\mu$ and dark vector $V_\mu$ fields, $\kappa$ is the kinetic mixing angle, and $J_{em}^\mu$ is the electromagnetic current.

The DM absorption rate is 
\be
R =\frac{1}{\rho}\frac{\rho_{DM}}{m_{DM}}\langle n \sigma_{abs}v\rangle \,,
\ee
where $\rho$ is the density of the crystal, $\rho_{DM}\sim 0.3\,$GeV/cm$^3$ is the local DM mass density, $m_{DM}=m_V$ for the dark photon model of eq.~\eqref{eq:lagrang}, $n$ is the number density of the molecules in the crystal, $\sigma_{abs}=\sigma_{abs}(m_V)$ is the DM--molecule absorption cross section, and the local DM velocity $v\sim10^{-3}$. We follow Refs.~\cite{An:2013yfc,An:2014twa} in relating the dark vector absorption cross section to that of photons, taking into account in-medium effects, 
\be
\langle n \,\sigma_{abs}(m_V)\,v\rangle =  \kappa_{eff}^2 \langle n\, \sigma_{abs}^{\gamma}(\omega=m_V)\,c\rangle \,, 
\label{eq:xsectogamma}
\ee
where $\sigma_{abs}^{\gamma}(\omega=m_V)$ is the EM photon--molecule absorption cross section for photons of energy $\omega$, and where for clarity we restored a factor of $c$. 
The effective in-medium mixing angle is defined as,
\be
\kappa_{eff}^2 \simeq \kappa^2 \frac{1}{|\epsilon(\omega)|^2} \,,
\label{eq:inmed}
\ee
where $\epsilon$ is the relative permittivity of the SMM. 
%

There is currently little available detailed data on the IR absorption spectrum of the majority of known SMM crystals. To obtain a representative example of the sensitivity to the dark vector model we use data taken on the first discovered and most widely studied SMM, Mn$_{12}$O$_{12}$-acetate (Mn$_{12}$-acetate). Although this SMM does not have the desired tuning parameters to be used as a DM detector, the absorption of EM radiation of energy $\gtrsim$\,meV that we consider is not sensitive to such parameters that are properties of the entire molecule.  (Absorption of light of frequency Hz-MHz is dependent on such properties, see {\it e.g.} Ref.~\cite{PhysRevB.59.11147}.) 
We proceed using the approximation $\kappa\simeq\kappa_{eff}$, and defer a more detailed study to when crystals specific to a (prototype) DM detector have been identified. However, because SMMs are insulators, we expect that this is a reasonable approximation ({\it i.e.} the sensitivity to $\kappa$ is within one or two orders of magnitude of that of $\kappa_{eff}$ across the entire energy range we consider---we explicitly verified this in our representative example below in regions of energy where such data was available.). 

In the meV--100\,meV regime, photons are  absorbed by the vibrational modes between ions in the molecule ({\it e.g.} Mn-O, C=O, {\it etc.})---while these give rise to spectral `fingerprints' that differ depending on the composition of the SMM, the broad features are similar. For instance any molecule with a C=O bond will show an absorption feature in the 190--220 meV region. Similarly, the Mn-O bond is of course absent in SMMs based on different metallic ions, but will be replaced by {\it e.g.} a Fe-O, Fe-N, Dy-O {\it etc.} bond that absorbs at a different frequency.  

In the 100\,meV--10\,eV regime, EM absorption takes place via a restructuring of electronic configurations in the molecule. The spectrum for this compound has been found to be diffuse~\cite{PhysRevB.65.054419} (in line with theoretical predictions \cite{PhysRevB.59.6927, PhysRevB.60.9566}),with very broad excitations identified {\it e.g.} Mn inner $\to$ outer and  O $p$ $\to$ Mn $d$ charge transfers.

For the meV--100\,meV regime we use data from Ref.~\cite{PhysRevB.63.214408}, and for 100\,meV--10\,eV we use data from Ref.~\cite{PhysRevB.65.054419}; from this we are able to obtain $\sigma_{abs}^\gamma$ for Mn$_{12}$-acetate. See App.~\ref{app:a} for further details.

\begin{figure}
\begin{center}
\includegraphics[width=13cm]{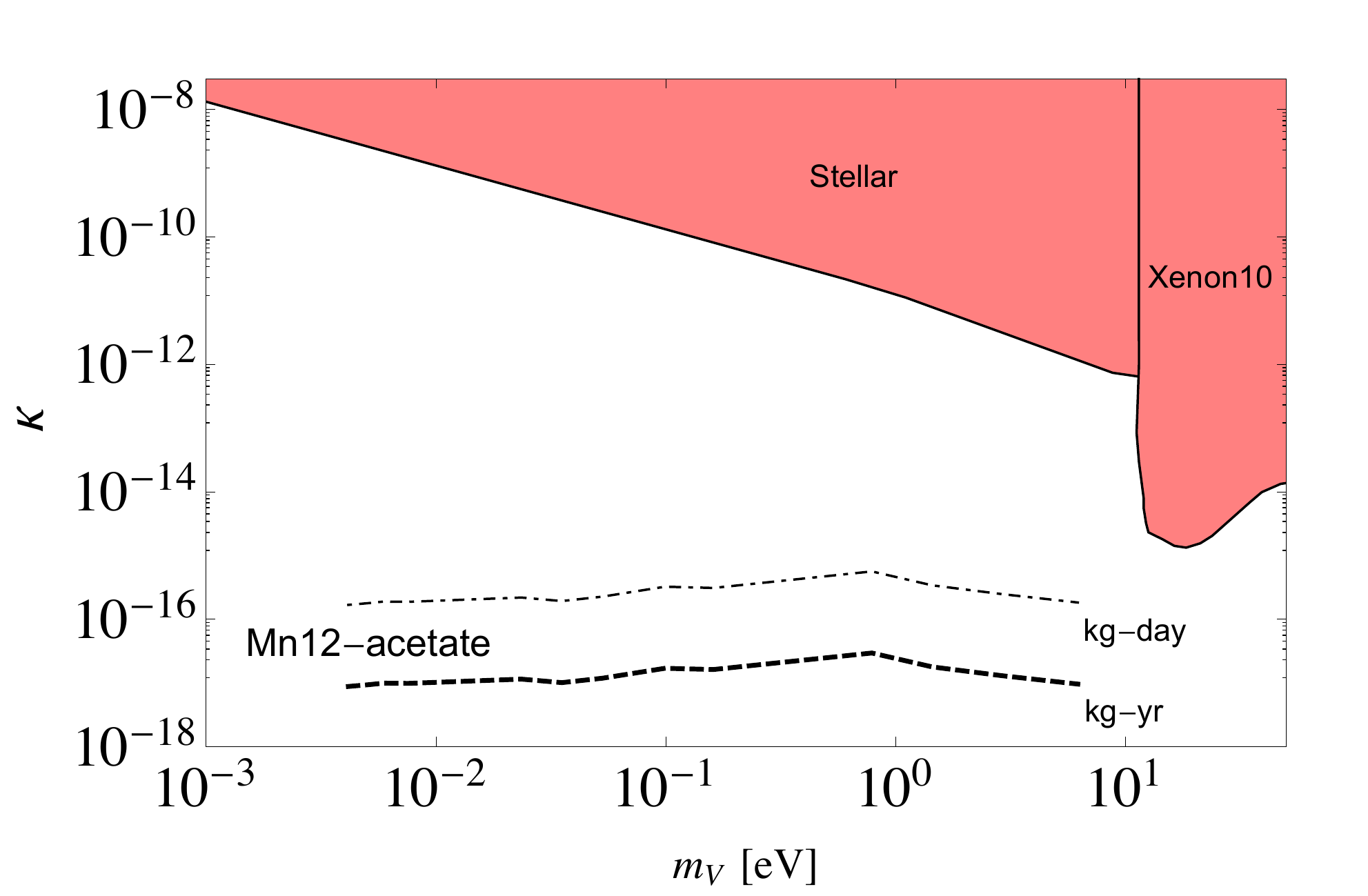}
\caption{Estimated sensitivity to absorption of dark vector DM in Mn$_{12}$-acetate, assuming an aggressive sensitivity of 1 event/kg year (dashed), and a sensitivity of 1 event/kg day (dot-dashed). The absorption data from Refs.~\cite{PhysRevB.65.054419},~\cite{PhysRevB.63.214408} (described in the appendix) has been smoothed,  an interpolation used in the region $m_V\sim0.2-0.5$\,eV, for which no data was available, and we use the approximation $\kappa\simeq \kappa_{eff}$ (see text).}
\label{fig:mn12acdata}
\end{center}
\end{figure}

The above absorption mechanisms that we are exploiting are localised---they are {\it intra}-molecular excitations. This is very different to excitations of phonon modes that are coherent across the entire crystal; such modes require long-range order. In principle, {\it inter}-molecular excitations would lead to such phonon modes (at lower energies); however, crystal defects and impurities change the phonon spectrum significantly. In general, we expect that such disorder could be advantageous, as localisation typically leads to larger absorption (or scattering) cross sections through making many more phonon modes available for radiation to couple to (also, these phonon modes typically give zero-order EM dipoles). Describing such processes theoretically, however, is more difficult than in a very pure system, and we postpone a detailed discussion to future work; for the present we use the available data for the localised absorption processes that are active at the energies considered above.

Using the absorption data for Mn$_{12}$-acetate, we plot the expected sensitivity to the dark vector model in Fig.~\ref{fig:mn12acdata}, assuming an aggressive sensitivity of 1 event per kg-year, and a SMM energy threshold below that of the DM mass. The neutrino background in the $1-100$\,meV range of energy is approximately 1 / kg-year \cite{Strigari:2009bq}, and as such the curves rest upon the neutrino floor; above this energy the curve is above the floor. Also shown are existing exclusion limits on the model of eq.~\eqref{eq:lagrang}, from stellar constraints \cite{An:2013yfc} and the Xenon10 experiment \cite{An:2014twa}. 
The only limits that exist in the region where we plot SMM projections are set by stellar bounds. (See also Refs.~\cite{Hochberg:2016sqx} and~\cite{Hochberg:2016ajh} for estimated sensitivities from proposed semiconductor and superconductor detectors absorbing in this energy regime.)
 Our results suggest that an improvement in sensitivity over stellar bounds across the entire energy range is possible, with an improvement of several orders of magnitude in the meV$-$10\,meV region. This would significantly extend the reach of DM direct detection experiments below the current lower bound on mass.

\section{Conclusions}
\label{sec:conc}

We have discussed the potential of single molecule magnets for use in the direct detection of DM, with energy thresholds down to $\sim$\,meV. These crystals possess a built-in amplification mechanism---magnetic deflagration, or a spin avalanche---a phenomenon that is tunable, such that a detector concept that operates in a similar fashion to a conventional bubble chamber can be envisaged.

The mechanism that sets off the spin deflagration is rather general---any process that deposits enough localised heat in the crystal will trigger it. The energy gap, or threshold, for this is one of the tunable aspects of our concept. 
Residual radioactive backgrounds need to be studied, but there is hope that they can be managed to an acceptable level. We mention here that a prototype experiment running with a higher threshold of $\sim$10\,eV per $\sim$nm$^3$ region would not be triggered by electron backgrounds, since the energy loss of an electron as it passes through the detector falls below this threshold (see {\it e.g.}~\cite{turner2008atoms}).

Although there is a lack of energy reconstruction for a given scattering/absorption process, in a discovery scenario (where DM  interacts at a reasonable rate) the energy threshold of the SMM could be tuned---one simple and smooth way would be by varying the magnitude of the external field---and in this way  information on the mass scale of the DM could in fact be reconstructed.

Importantly, the spin deflagration mechanism does not require a pure sample---impurities and defects (disorder) do not need to be eliminated. This touches on a perhaps more important general observation: disorder can be tolerable, and even advantageous, since it typically causes localisation that brings about larger interaction cross sections.

The outcome of the current work suggests it would be good to study more general glassy materials, {\it e.g.} spin glasses, to determine which possess the most favourable DM detector qualities.  For instance, in the SMM crystals that we have surveyed, we need temperatures of $\sim 0.1$\,K in order to make the meta-stable state `stable' on a detector run timescale. It is possible that other frustrated systems could realize these conditions at higher temperatures, which could significantly improve the experimental condition.

We used readily available IR absorption data to estimate the projected sensitivity to a dark vector model of DM. We found a sensitivity extending three orders of magnitude in mass below that of existing experiments, and several orders of magnitude below that of existing astrophysical bounds. To fully explore the sub-GeV DM parameter space, including scalar and fermionic DM, other triggering mechanisms should be investigated ({\it e.g.} Raman scattering, scattering in general). 

Given the relatively low cost and ease of growing SMM crystals, and the prospects demonstrated in this study, we believe there is a strong case to begin to experimentally examine their use as DM detectors.

\acknowledgements
We would like to thank the Heising-Simons Foundation for organizing a workshop at UC Berkeley during which this idea was conceived, and Simon Knapen, Steve Liddle, Tongyan Lin, Dan McKinsey and Matt Pyle for useful discussions. PB is supported by U.S. NSF grant CHE-1464841; GG is supported, in part, by U.S. DOE grant DE-SC0009841; TM is supported by U.S. DOE grant DE-AC02-05CH11231, and by WPI, MEXT, Japan; SR was supported in part by the NSF under grants PHY-1417295 and PHY-1507160, the Simons Foundation Award 378243 and  the Heising-Simons Foundation grant 2015-038.

\appendix

\section{Absorption in Mn$_{12}$-acetate}
\label{app:a}

\begin{figure}
\begin{center}
\includegraphics[width=10cm]{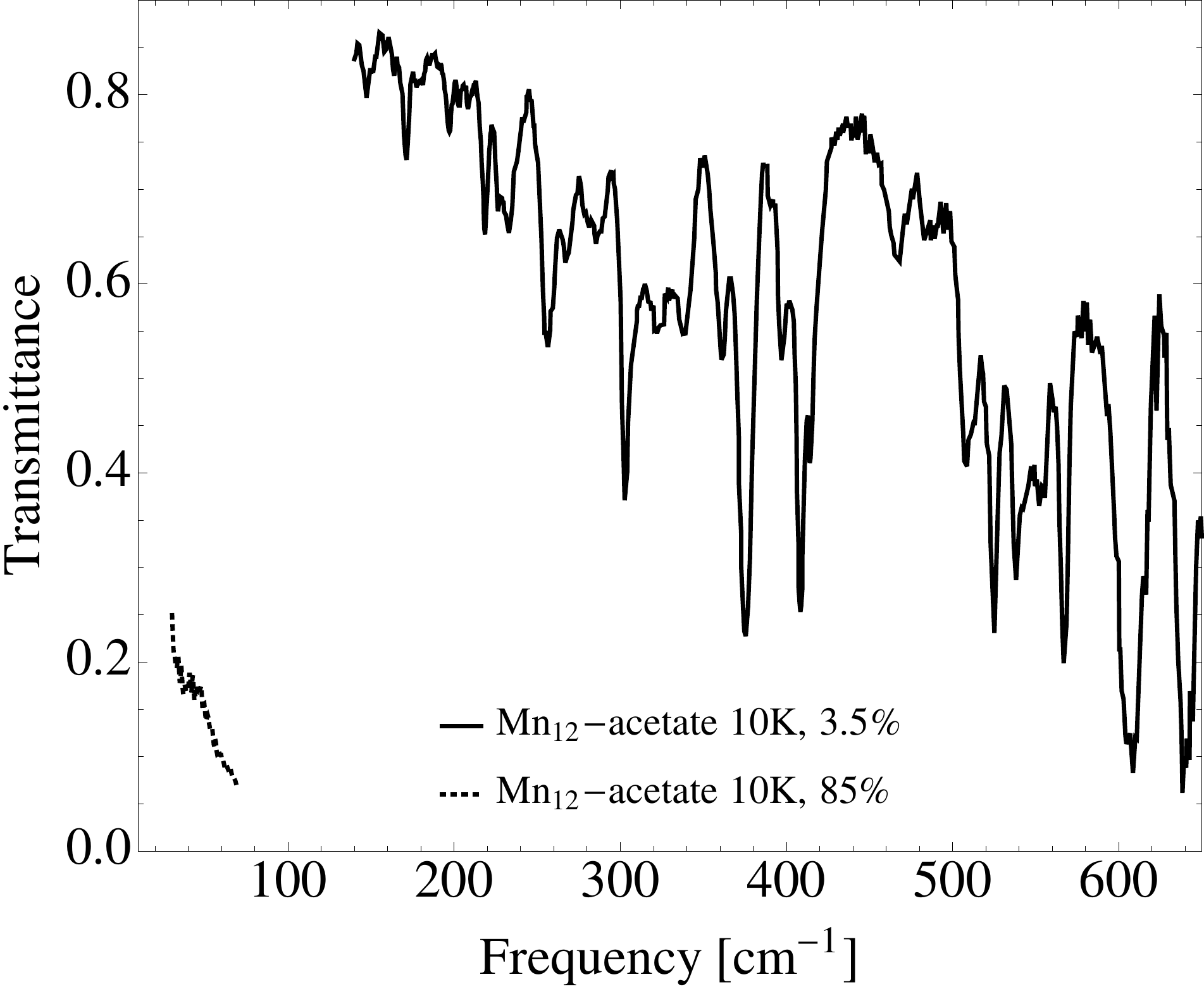}
\caption{Transmittance data for a Mn12-acetate power, reproduced from Ref.~\cite{PhysRevB.63.214408}. }
\label{fig:mn12actrans}
\end{center}
\end{figure}

\begin{figure}
\begin{center}
\includegraphics[width=10cm]{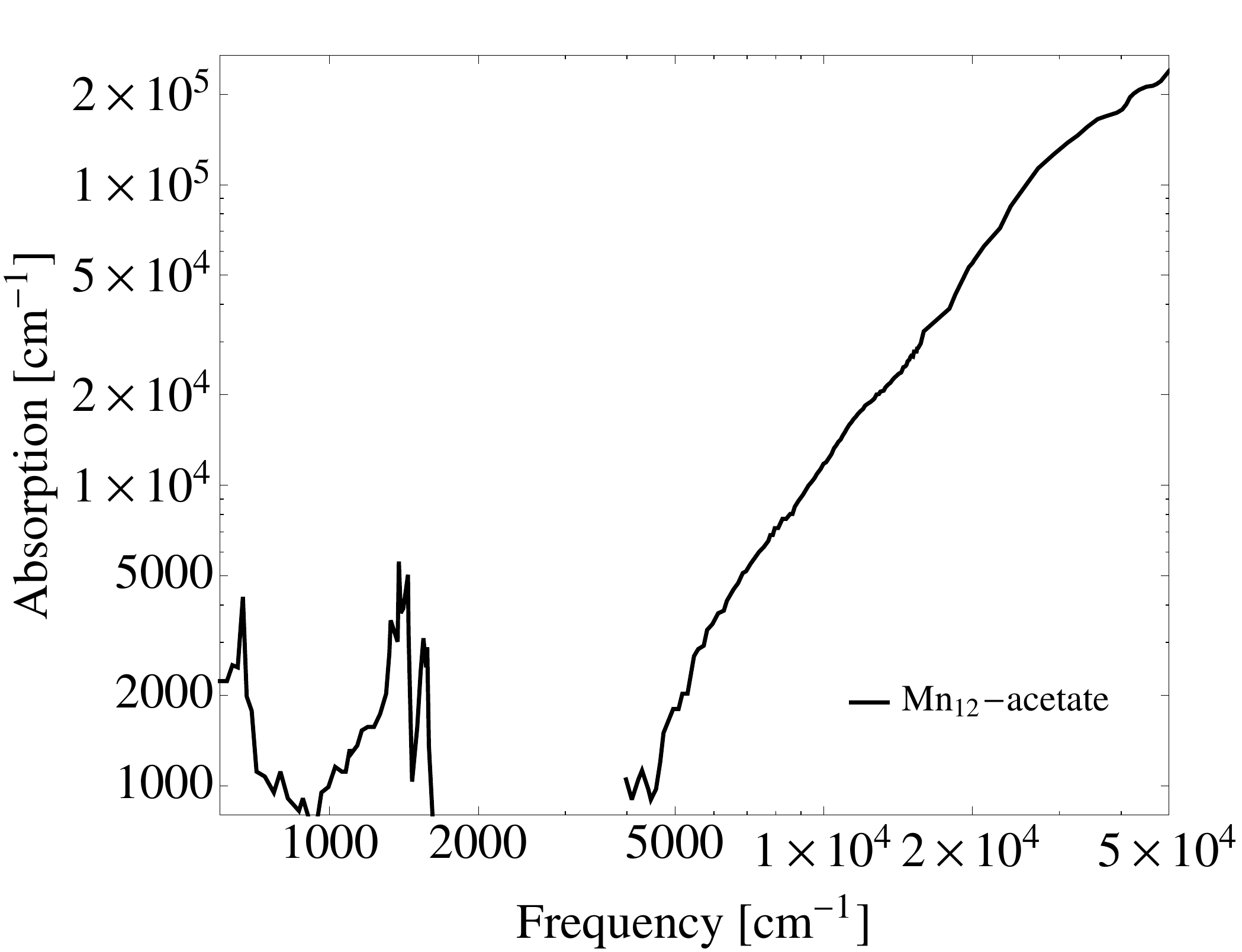}
\caption{Absorption spectrum for Mn12-acetate, reproduced from Ref.~\cite{PhysRevB.65.054419}.}
\label{fig:mn12acabs}
\end{center}
\end{figure}

In Fig.~\ref{fig:mn12actrans} we reproduce the transmittance, $T$, for Mn$_{12}$-acetate which was presented in Ref.~\cite{PhysRevB.63.214408}, in the frequency ranges $30-70$\,cm$^{-1}$ and $140-650$\,cm$^{-1}$ . This data is taken at zero magnetic field and at $10$\,K. In this experiment Mn$_{12}$-acetate was finely ground and distributed in a paraffin pellet, of thickness $l$, at a concentration $h$ ($h=0.85$ and $h=0.035$ for the two different curves shown in Fig.~\ref{fig:mn12actrans}). The spectral lines in this frequency range are largely identified with Mn-O bond vibrations.  While pronounced trough positions are reasonably stable under temperature (the range 10\,K-100\,K was studied), they sharpen slightly at lower temperatures. In an external magnetic field, as would be the case in the running of our proposed DM detector, these troughs broaden and shift to a small degree.

If reflection is neglected, the absorption coefficient, $\alpha(\omega)$, can be obtained from the transmittance via the Beer-Lambert law,
\be
T(\omega) = \exp(-\alpha(\omega) \,l\,h).
\label{eq:beer}
\ee
Reflection measurements were not reported in \cite{PhysRevB.63.214408}, and in fact the authors of that work calculate relative absorption using eq.~\eqref{eq:beer}. Thus we infer that the reflectance in the experimental setup was small (this is plausible given the SMM was finely ground and distributed in paraffin), such that eq.~\eqref{eq:beer} is a good approximation. The thickness, $l$, was also not reported; we fix this parameter by matching onto the calculated absorption spectrum at higher energies, presented below. 

Ref.~\cite{PhysRevB.65.054419} studies both transmittance and reflectance of Mn$_{12}$-acetate crystals over the frequency range 600-50\,000\,cm$^{-1}$. From this the authors calculate the absorption spectrum for light incident both parallel and perpendicular to the magnetic anisotropy axis; we reproduce the average in Fig.~\ref{fig:mn12acabs}. In the low energy data sample, spectral peaks at 600\,cm$^{-1}$ are again observed and correspond to vibrations of the Mn$_{12}$ crown; spectral features around $\sim1000-1500$\,cm$^{-1}$ are identified in Ref.~\cite{PhysRevB.66.174437} as Mn-acetate, C-O, and acetate stretching in Raman scattering data (in  \cite{PhysRevB.66.174437}, similar `ligand' resonances in IR absorbtion of the SMM Fe$_8$Br$_8$~\cite{ANIEBACK:ANIE198400771} were also identified). Several higher energy, very broad spectral features are identified in \cite{PhysRevB.65.054419}, {\it e.g.} a density of states effect related to a charge transfer between Mn ions at 15\,900\,cm$^{-1}$, and a charge transfer from O to Mn ions 36\,600 and 42\,350\,cm$^{-1}$---these are not easily seen in Fig.~\ref{fig:mn12acabs}.

The EM absorption cross section is directly related to the absorption coefficient, 
\be
\sigma_{abs}^\gamma(\omega)=\alpha(\omega)\,n \,,
\ee
where $n$ is the number density of molecules.

\bibliographystyle{apsrev4-1}
\bibliography{bibliography.bib}

\end{document}